\documentclass[aps,12pt, a4paper]{revtex4}

\usepackage[utf8]{inputenc}
\usepackage[T1]{fontenc}
\usepackage{amsmath}
\usepackage{amsfonts}
\usepackage{amssymb}
\usepackage{graphics,graphicx}

\usepackage{graphicx,color}
\usepackage{subfigure} 
\usepackage{wrapfig} 
\usepackage{epstopdf}
\graphicspath{{figuras/}}

\usepackage[breaklinks=true]{hyperref}
\usepackage{setspace}

\newcommand{\bes}{\begin{subequations}}
\newcommand{\ees}{\end{subequations}}
\def\ben{\begin{eqnarray}}
\def\een{\end{eqnarray}}
\def\be{\begin{equation}}
\def\ee{\end{equation}}

\begin{document}

\title{Rotating black hole with a probe string in Horndeski Gravity}
\author{F. F. Santos$^{1}$}
\email{fabiano.ffs23@gmail.com}   
\address{$^{1}$Departamento de F\'\i sica, Universidade Federal da Para\'iba, Caixa Postal 5008, 58051-970, Jo\~ ao Pessoa, Para\' iba, Brazil.}

\begin{abstract}
In this work, we present the effect of a probe string on the complexity of a black hole according to the CA (Complexity equals action) conjecture on Horndeski's gravity. In our system, we consider a particle moving on the boundary of black hole spacetime in ($2+1$)-dimensions. To obtain a dual description, we need to insertion a fundamental string on the bulk spacetime. The effect of this string is given by the Nambu-Goto term. For the Nambu-Goto term, we can analyze the time development of this system, which is affected by the parameters of Horndeski's gravity. In our case, we show some interesting complexity properties for this gravity. 
\end{abstract}

\maketitle
\section{Introduction}
\label{intro}
Over the past few years, General Relativity has been supported by substantial evidence and observations, which have been witnessed in many astrophysical scenarios, these in turn range from Eddington’s measurement of light deflection in 1919 to the recent direct observation of gravitational waves by collaboration LIGO \cite{Dyson:1920cwa,Abbott:2016blz}. These observational data served as motivation to consider the modified gravity as a theory that has given rich support to phenomenology \cite{Casalino:2018mna}. However, some problems considered fundamental to be understood in the context of General Relativity, such as dark matter, dark energy, and the inflationary phase of the Universe. In recent years, modifications of General Relativity have been proposed; however, to carry out such modifications of General Relativity, it is necessary to maintain some of its essential properties, which are second-order equations of motion resulting from an invariable action by diffeomorphism and Lorentz’s invariance \cite{Heisenberg:2018vsk}. By maintaining these properties, additional degrees of freedom of propagation can be added in the gravity sector that is consistent to include additional fields such as scalars, vectors, or tensors. However, to deal with these problems in Einstein’s gravity, one way was to couple the theory with scalar fields. In this sense, these efforts led to the development of the now known Galileo theories, which are theories of scalar tensors \cite{Nicolis:2008in}. Besides, these studies led to the rediscovery of Horndeski’s gravity \cite{Horndeski:1974wa}. This theory was presented in 1974 by Horndeski for further discussions see \cite{Horndeski:1974wa,Bruneton:2012zk,Brito:2018pwe,Santos:2019ljs}, it is characterized by being a theory of scalar tensors with second-order field equations and second-order energy-moment tensor \cite{Heisenberg:2018vsk,Horndeski:1974wa,Cisterna:2014nua,Anabalon:2013oea,Bravo-Gaete:2013dca,Rinaldi:2012vy}. The Lagrangian for this theory produces second-order equations of motion \cite{Cisterna:2014nua,Anabalon:2013oea,Deffayet:2011gz,VanAcoleyen:2011mj,Gomes:2015dhl,Rinaldi:2016oqp,Cisterna:2017jmv,Zumalacarregui:2013pma,Feng:2015oea,Brito:2019ose,Cisterna:2016vdx}. This theory also includes four arbitrary functions of the scalar field and its kinetic term.

In recent years, the Horndeski gravity \cite{Cisterna:2016vdx} has been shown to support static black hole solutions with asymptotically anti-de Sitter behavior. In general, on the astrophysical scale, black holes are not expected to be static and has spherical symmetry, but asymmetric because they have angular momentum \cite{Tattersall:2018nve}. However, asymmetric black holes are the objects expected to be involved in the events detected in the gravitational wave observatories by the collaborations LIGO and VIRGO \cite{Abbott:2016blz,Abbott:2016nmj,Abbott:2017vtc}. As discussed by \cite{Tattersall:2018nve}, the author performed an extension of the calculation of perturbations from a Schwarzschild black hole to a Kerr black hole with a slow rotation, showing an agreement of the almost normal modes analytically with the calculated numerical frequencies. However, the ’bald’ black holes in Horndeski’s gravity are identical to their equivalents in General Relativity. They can display a modified gravitational wave signal during the ringdown, which is characterized by only one or two observation parameters useful for attempts to restrict gravity in this new era of gravitational-wave astronomy. 

In addition to the results explored by \cite{Tattersall:2018nve} to explore the equivalence between ’bald’ black holes in Horndeski’s gravity and General Relativity, another scenario explored by \cite{Bravo-Gaete:2014haa} showed that due to a particular truncation of Horndeski’s action, it reduces to EinsteinHilbert’s Lagrangian with a cosmological constant and a scalar field, which has a dynamics governed by their usual kinetic term, together with a non-minimal kinetic coupling. In this sense, the radial component of the conserved current has to disappear, providing a solution with a geometry of a BTZ black hole with a radial scalar field that is well defined on the horizon.

Slowly rotating black holes in Horndeski’s gravity context has been presented by \cite{Maselli:2015yva}, where first-order rotational corrections were considered for a wide range of black hole solutions in Horndeski’s gravity. It has been shown that the drag function, which describes the rotational corrections of leadership order, is precisely the same as in General Relativity for all known black hole solutions by Horndeski. However, a very relevant fact for these black holes with slow rotation in Horndeski’s gravity is that the no-hair theorem is valid in the static regime and can be extended to apply also to slow rotation black holes \cite{Maselli:2015yva,Hui:2012qt,Sotiriou:2015pka}. In this work, we will consider the computational complexity of a rotating black hole in (2+1)-spacetime in Horndeski’s gravity. For further discussions regarding the computational complexity of black holes, see the following references \cite{Nagasaki:2017kqe,Nagasaki:2018csh,deBoer:2017xdk,Banerjee:2019vff,Brown:2015lvg,Brown:2015bva}. For this black hole, butterfly effects caused by a small disturbance in an asymptotic region of this black hole were presented by \cite{Reynolds:2016pmi}. However, in our study, we considered the effect of the string moving on this spacetime geometry of the BTZ black hole. This work is summarized as follows: In Sec.~\ref{v1}, we present the Horndeski gravity. In Sec.~\ref{v2}, we address the issue of finding black hole ansatz, and we explore the effect of the probe string in the black hole. In the Sec.~\ref{v3}, we investigate the Nambu-Goto (NG) action of the string moving in the black hole. Finally, in Sec.~\ref{v4}, we present our conclusions.
\section{Horndeski gravity}\label{v1}
In this section, we present the model of John Lagrangian \cite{Bruneton:2012zk}, which is a specific model related to F4 theories for further discussions on F4 theories see \cite{Charmousis:2011bf,Charmousis:2011ea}, they are a particular subclass of Horndeski's theory.
This subclass has attracted attention in recent years when carrying out investigations around the standard kinetic term for the scalar \cite{Starobinsky:2016kua}, providing a model that is sometimes called Fab Five (F5), which gained high visibility in the study of the cosmological scenarios to know \cite{Santos:2019ljs}. The following action gives John Lagrangian \cite{Bruneton:2012zk} model
\begin{equation}
S[g_{\mu\nu},\phi]=\int{d^{4}x\sqrt{-g}\left[\kappa(R-2\Lambda)-\frac{1}{2}(\alpha g_{\mu\nu}-\gamma G_{\mu\nu})\nabla^{\mu}\phi\nabla^{\nu}\phi\right]}+S_{m}.\label{1}
\end{equation}
Where $\kappa=(16\pi G)^{-1}$, we can define a new field $\phi^{'}\equiv\psi$, and $S_{m}$ describes an ordinary matter that is supposed to be a perfect fluid. We can see that in action the field has a dimension of $(mass)^{2}$ with the parameters $\alpha$ and $\gamma$ controlling the strength of the kinetic couplings, note that $\alpha$ is dimensionless and $\gamma$ has a dimension of $(mass)^{-2}$. The Einstein-Horndeski field equations for the action (\ref{1}) can be written formally by varying this action, that is, $\delta S[g_{\mu\nu},\phi]$ and assuming $S_{m}=$constant, we can write the equations of motion as in the form
\begin{eqnarray}
E_{\mu\nu}[g_{\mu\nu},\phi]&=&G_{\mu\nu}+\Lambda g_{\mu\nu}-\frac{\alpha}{2\kappa}\left(\nabla_{\mu}\phi\nabla_{\nu}\phi-\frac{1}{2}g_{\mu\nu}\nabla_{\lambda}\phi\nabla^{\lambda}\phi\right)\label{2}\\
                  &-&\frac{\gamma}{2\kappa}\left(\frac{1}{2}\nabla_{\mu}\phi\nabla_{\nu}\phi R-2\nabla_{\lambda}\phi\nabla_{(\mu}\phi R^{\lambda}_{\nu)}-\nabla^{\lambda}\phi\nabla^{\rho}\phi R_{\mu\lambda\nu\rho}\right)\nonumber\\
									&-&\frac{\gamma}{2\kappa}\left(-(\nabla_{\mu}\nabla^{\lambda}\phi)(\nabla_{\nu}\nabla_{\lambda}\phi)+(\nabla_{\mu}\nabla_{\nu}\phi)\Box\phi+\frac{1}{2}G_{\mu\nu}(\nabla\phi)^{2}\right)\nonumber\\
									&-&\frac{\gamma}{2\kappa}\left[-g_{\mu\nu}\left(-\frac{1}{2}(\nabla^{\lambda}\nabla^{\rho}\phi)(\nabla_{\lambda}\nabla_{\rho}\phi)+\frac{1}{2}(\Box\phi)^{2}-(\nabla_{\lambda}\phi\nabla_{\rho}\phi)R^{\lambda\rho}\right)\right],\nonumber\\
E_{\phi}[g_{\mu\nu},\phi]&=&\nabla_{\mu}J^{\mu};\quad J^{\mu}=[(\alpha g^{\mu\nu}-\gamma G^{\mu\nu})\nabla_{\nu}\phi].\label{3}
\end{eqnarray}
Using the fact that $E_{\mu\nu}[g_{\mu\nu},\phi]=0$ and $E_{\phi}[g_{\mu\nu},\phi]=0$, we can write
\begin{equation}
G_{\mu\nu}+\Lambda g_{\mu\nu}=\frac{1}{2\kappa}T_{\mu\nu},\label{4}
\end{equation}
where $T_{\mu\nu}=\alpha T^{(1)}_{\mu\nu}+\gamma T^{(2)}_{\mu\nu}$ and the energy-momentum tensors $T^{(1)}_{\mu\nu}$ and $T^{(2)}_{\mu\nu}$ take the following format
\begin{equation}\begin{array}{rclrcl}
T^{(1)}_{\mu\nu}&=&\nabla_{\mu}\phi\nabla_{\nu}\phi-\frac{1}{2}g_{\mu\nu}\nabla_{\lambda}\phi\nabla^{\lambda}\phi\\
T^{(2)}_{\mu\nu}&=&\frac{1}{2}\nabla_{\mu}\phi\nabla_{\nu}\phi R-2\nabla_{\lambda}\phi\nabla_{(\mu}\phi R^{\lambda}_{\nu)}-\nabla^{\lambda}\phi\nabla^{\rho}\phi R_{\mu\lambda\nu\rho}\\
              &-&(\nabla_{\mu}\nabla^{\lambda}\phi)(\nabla_{\nu}\nabla_{\lambda}\phi)+(\nabla_{\mu}\nabla_{\nu}\phi)\Box\phi+\frac{1}{2}G_{\mu\nu}(\nabla\phi)^{2}\\
							&-&g_{\mu\nu}\left[-\frac{1}{2}(\nabla^{\lambda}\nabla^{\rho}\phi)(\nabla_{\lambda}\nabla_{\rho}\phi)+\frac{1}{2}(\Box\phi)^{2}-(\nabla_{\lambda}\phi\nabla_{\rho}\phi)R^{\lambda\rho}\right].\label{5}
\end{array}\end{equation} 
And the scalar field equation is given by
\begin{equation}
\nabla_{\mu}[(\alpha g^{\mu\nu}-\gamma G^{\mu\nu})\nabla_{\nu}\phi]=0.\label{6}
\end{equation}

\section{Black hole solutions and probe string}\label{v2}

Let us consider for Horndeski's gravity that the string is moving in the spacetime of the BTZ black hole \cite{Reynolds:2016pmi}. The BTZ black hole at ($2+1$)-dimensions is given by
\begin{equation}
ds^{2}=-f(r)dt^{2}+r^{2}\left(d\chi-\frac{J}{r^{2}}dt\right)^{2}+\frac{dr^{2}}{f(r)}.\label{7}
\end{equation}
where $J$ is the angular momentum, to escape the no-hair theorem, which has been well discussed in \cite{Bravo-Gaete:2013dca}, we impose that the conserved current's radial component disappears identically without restricting the radial dependence of the scalar field:
\begin{equation}
\alpha g_{rr}-\gamma G_{rr}=0\label{8}.
\end{equation}
Mind that $\phi^{'}(r)\equiv\psi(r)$, we can easily see that this condition annihilates $\psi^2(r)$, regardless of its behavior on the horizon. It is possible to find the function $f(r)$ using the equation (\ref{8}). Thus, equation (\ref{6}) is satisfied with the following solution
\begin{eqnarray}
f(r)&=&-M+\frac{\alpha r^{2}}{\gamma}+\frac{J^{2}}{r^{2}},\label{9}\\
\psi^{2}(r)&=&-\frac{2\kappa(\alpha+\gamma\Lambda)}{\alpha\gamma f(r)}.\label{10}
\end{eqnarray}
The Einstein-Horndeski field equations (\ref{4}) and (\ref{6}) are satisfied by these equations. Besides, as discussed by \cite{Anabalon:2013oea}, we have that $\alpha/(\gamma)=l^{-2}_{AdS}$ is defined as an effective radius of AdS $l_{AdS}$. For the case in which we have limitations in the storage of information, we have that the area of the black hole delimits the information \cite{Brown:2015lvg,Brown:2015bva}. The black hole entropy, which can be obtained by applying the first law of black hole thermodynamics $dM=TdS$, can be written as
\begin{eqnarray}
&&S=\frac{1}{2G}\int^{r_{h}}_{0}{\frac{1}{T(r_{h})}\frac{dM}{dr_{h}}dr_{h}}=\frac{2\pi r_{h}}{G}=\frac{A}{4G}\label{10.1}\\
&&M(r_{h})=\left(\frac{\alpha r^{2}_{h}}{\gamma}+\frac{J^{2}}{r^{2}_{h}}\right)\label{10.2}\\
&&T(r_{h})=\frac{1}{2\pi}\left(\frac{\alpha r_{h}}{\gamma}-\frac{J^{2}}{r^{3}_{h}}\right)\label{10.3}
\end{eqnarray}
Which obeys the celebrated Hawking's law. If an object can be forced to undergo a gravitational collapse by adding mass, the second law of thermodynamics insists that it must have less entropy than the resulting black hole. Now, we will present the effect of the string moving in this spacetime geometry. For this, we will calculate the induced metric using the parameters $\tau$ and $\sigma$ in the world-sheet of the fundamental string. These parameters are given as follows:
\begin{eqnarray}
t=\tau,\quad r=\sigma\quad,\chi=v\tau+\xi(\sigma),\label{11}
\end{eqnarray} 
Where $v$ is a constant velocity, and $\xi(\sigma)$ is a function that determines string's shape. However, the metric induced in the world-sheet is given by
\begin{eqnarray}
&&ds^{2}_{ind}=H(\sigma)d\tau^{2}+G(\sigma)d\sigma^{2}+2F(\sigma)d\tau d\sigma\label{12}\\
&&H(\sigma)=-f(\sigma)+\left(v\sigma-\frac{J}{\sigma}\right)^{2}\nonumber\\
&&G(\sigma)=\frac{1}{f(\sigma)}+\sigma^{2}\xi^{'2}(\sigma)\nonumber\\
&&F(\sigma)=\xi^{'}(\sigma)(v\sigma^{2}-J)\nonumber
\end{eqnarray}
Through the equation (\ref{8}) we can provide that the solution for (\ref{12}), which is given by
\begin{eqnarray}
&&f(\sigma)=\left(v\sigma-\frac{J}{\sigma}\right)^{2}\label{13}\\
&&\psi^{2}(\sigma)=\frac{4\kappa\Lambda G(\sigma)(-F^{2}(\sigma)+G(\sigma)H(\sigma))}{\alpha(-2F^{2}(\sigma)+G(\sigma)H(\sigma))}\label{13.1}
\end{eqnarray}
Where this solution satisfies all equations (\ref{4}) and (\ref{6}). Now, integrating the equation (\ref{6}), we can write that
\begin{eqnarray}
\psi(\sigma)=\frac{-F^{2}(\sigma)+G(\sigma)H(\sigma)}{\alpha H(\sigma)}\label{13.2}
\end{eqnarray}

\section{Evaluating of the NG action}\label{v3}
We are now going to carry out the NG action evaluation process, which was recently presented by \cite{Nagasaki:2017kqe,Nagasaki:2018csh,deBoer:2017xdk,Banerjee:2019vff}. Let us address the action of the Nambu-Goto (NG) term, which is given by
\begin{eqnarray}
S_{NG}=-T_{s}\int{d\sigma^{2}\sqrt{-detg_{ind}}},\label{14}
\end{eqnarray}
where $T_{s}$ is the fundamental string tension and the horizon is determined by $f(r)=0$. Adding the Wilson loop implies the insertion of a fundamental string whose worldsheet has a limit in the Wilson loop. In this sense, we can calculate the NG action of this fundamental string. Besides, we can calculate the time derivative of the NG action, which is obtained by integrating the square root of the determinant of the induced metric, and write that
\begin{eqnarray}
\frac{dS_{NG}}{dt}=T_{s}\int^{r_{+}}_{r_{-}}{d\sigma\sqrt{\xi^{'2}(\sigma)(v\sigma^{2}-J)^{2}}},\label{15}
\end{eqnarray}
The Lagrangian is given by
\begin{eqnarray}
\mathcal{L}=T_{s}\xi^{'}(\sigma)(v\sigma^{2}-J),\label{16}
\end{eqnarray}
which has the following equations of motion
\begin{eqnarray}
\frac{d}{d\sigma}\frac{\partial\mathcal{L}}{\partial\xi^{'}(\sigma)}-\frac{\partial\mathcal{L}}{\partial\xi(\sigma)}=0\label{17}
\end{eqnarray}
Where through the equation of motion (\ref{17}), we can easily show that $v=0$, this fact characterizes a stationary string. So, by the equation (\ref{15}), we have
\begin{eqnarray}
\frac{dS_{NG}}{dt}=\left.T_{s}J\xi(\sigma)\right|^{r_{+}}_{r_{-}},\label{18}
\end{eqnarray}
We can see that through the equation (\ref{13}) combined with $H(\sigma)$, we can draw from the equation (\ref{13.2}) that $\xi(\sigma)=c_{\xi}/J$ where $c_{\xi}$ is an integration constant that we will take to be $c_{\xi}>0$ and $c_{\xi}<0$ in our analysis. So, we can write for equation (\ref{18}):
\begin{eqnarray}
\frac{dS_{NG}}{dt}=T_{s}c_{\xi}\left(\sqrt{\frac{\gamma M}{2\alpha}\left(1+\sqrt{1-\frac{4\alpha J^{2}}{\gamma M^{2}}}\right)}+\sqrt{\frac{\gamma M}{2\alpha}\left(1-\sqrt{1-\frac{4\alpha J^{2}}{\gamma M^{2}}}\right)}\right),\label{19}
\end{eqnarray}
 
\begin{figure}[!ht]
\begin{center}
\includegraphics[scale=0.7]{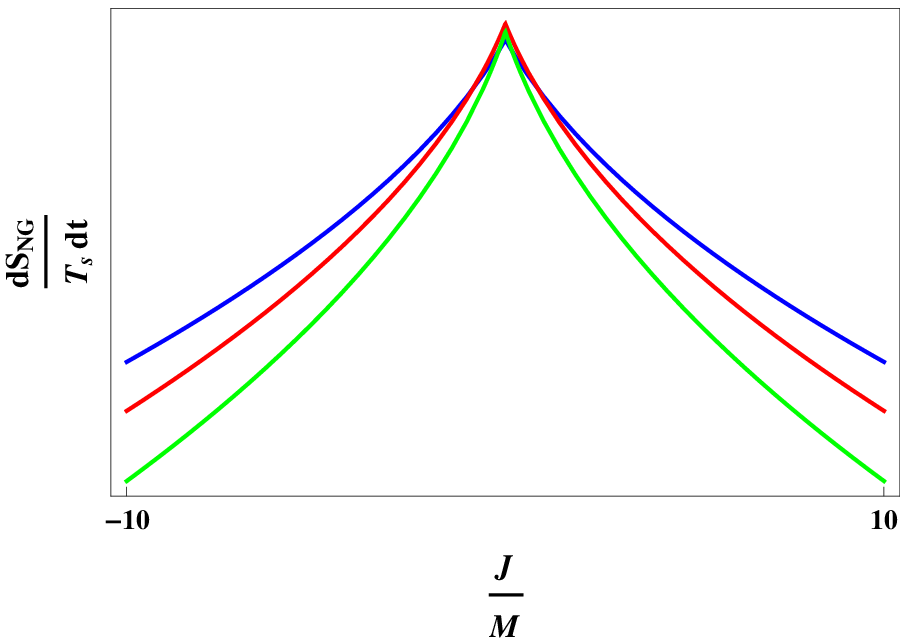}\hspace{1 cm}\includegraphics[scale=0.7]{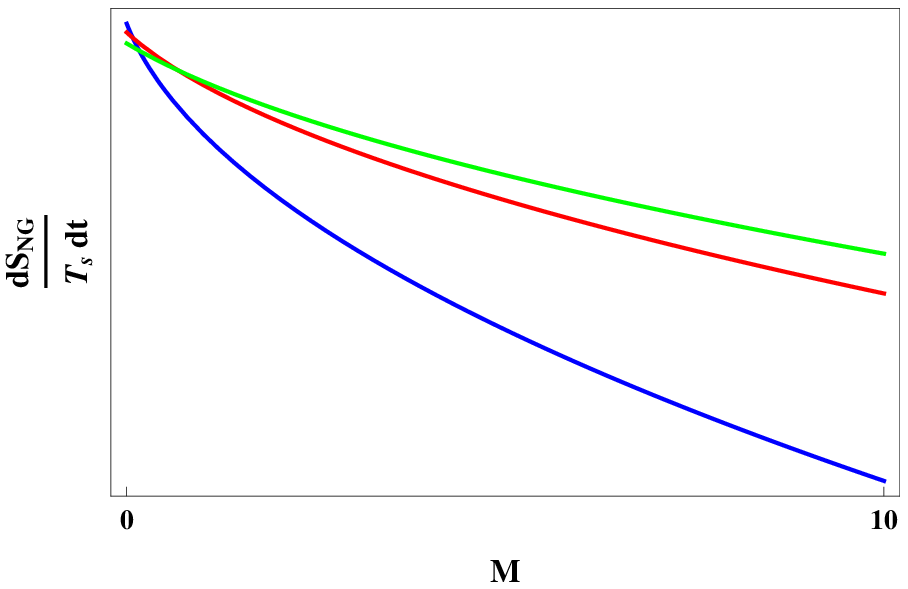}
\caption{BTZ: Action growth-Black hole angular momentum/Mass for $c_{\xi}=-1$ with the values $M=1$-$\alpha=0.1$-$\gamma=0.5$ (blue curve), $M=2$-$\alpha=0.5$-$\gamma=1$ (curve red), and $M=3$-$\alpha=1$-$\gamma=1.5$ (green curve). Action growth-Black hole mass (small mass region) for $q=-1$ with the values $J=0.2$-$\alpha=0.1$-$\gamma=0.5$ (blue curve), $J=0.4$-$\alpha=0.5$-$\gamma=1$ (curve red), and $J=0.6$-$\alpha=1$-$\gamma=1.5$ (green curve).}\label{p}
\label{planohwkhz}
\end{center}
\end{figure}

\begin{figure}[!ht]
\begin{center}
\includegraphics[scale=0.7]{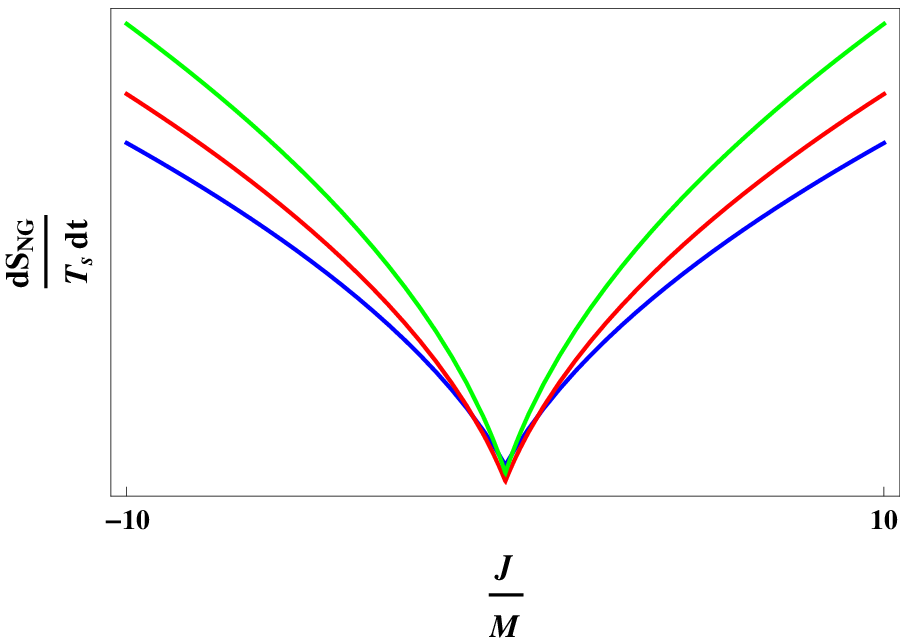}\hspace{1 cm}\includegraphics[scale=0.7]{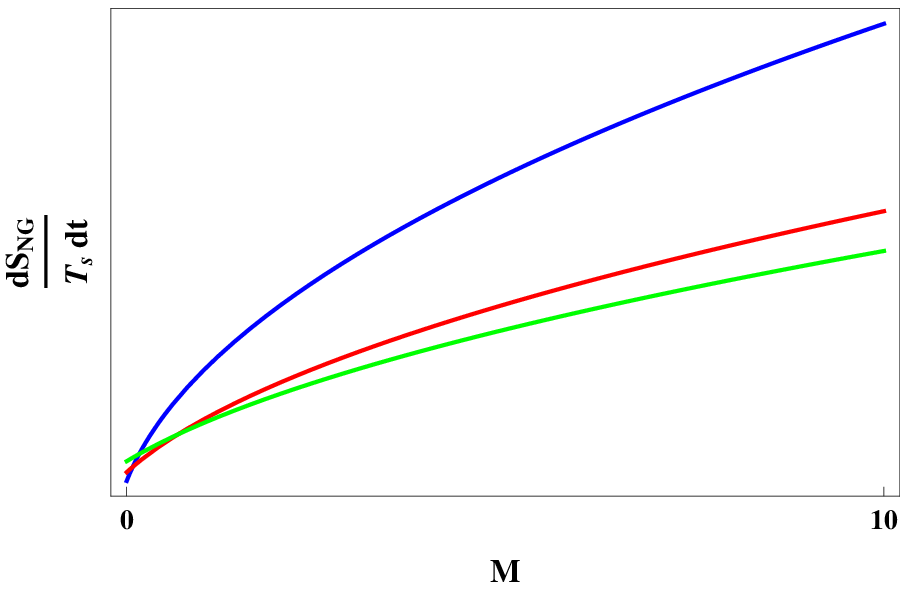}
\caption{BTZ: Action growth-Black hole angular momentum/Mass for $c_{\xi}=1$ with the values $M=1$-$\alpha=0.2$-$\gamma=0.5$ (blue curve), $M=2$-$\alpha=0.5$-$\gamma=1$ (curve red), and $M=3$-$\alpha=1$-$\gamma=1.5$ (green curve). Action growth-Black hole mass for $q=1$ with the values $J=0.2$-$\alpha=0.2$-$\gamma=0.5$ (blue curve), $J=0.4$-$\alpha=0.5$-$\gamma=1$ (curve red), and $J=0.6$-$\alpha=1$-$\gamma=1.5$ (green curve).}\label{w}
\label{planohwkhz}
\end{center}
\end{figure}
We can observe, according to figure \ref{p}, the dependence of the relationship between the growth of the action and the angular momentum/mass and mass of the black hole along with Horndeski's parameters. In this way, the increase in complexity reaches its maximum when the string is in a steady-state. Therefore, we can observe that the effect of complexity reaches its maximum when the relative speed is zero for a rotating black hole. As expected, this effect is greater for larger black holes, as in our case, where we consider the AdS$_{4}$ case. In figure \ref{w}, we have a dependence on the angular momentum/mass where we can see that the string rotates on a different axis, the relative speed never becomes zero. We can also notice that in figure \ref{w} the dependence of the growth of the action with the black hole mass, the parameters of Horndeski's theory also influence this behavior. When the angular momentum is small, we have a monotonically increasing function of the mass. As the angular momentum becomes larger, it stops increasing rapidly, and the extreme appears around $J=0.6$.
 
\section{Conclusion}\label{v4}

In this work, we show the effects of the probe string on a BTZ black hole, a solution of Horndeski's theory without any restrictions on the parameters. We have seen that although the string is stationary, it can provide maximum growth in complexity, which is obtained for $c_{\xi}<0$ along with the parameters of Horndeski's theory, see figure \ref{p}. However, for $c_{\xi}>0$ as shown in figure \ref{w} the growth of the action with the black hole mass, which has its behavior influenced by the parameters of Horndeski's theory where the complexity increases with the black hole mass giving us an idea of how complex the physical system. As we showed in the calculation of NG's action on the remaining flame, the peak is located at $v=0$, where this behavior derives from the expansion of time. We can perceive a decrease in complexity, as shown in figure \ref{p}, with the growth of the action with the the black hole mass, which is a monotonically decreasing function of the black hole mass. One might think that this is because complexity defines how complex the physical system is. So, if complexity decreases, we have less information from the physical system. So, this is a bigger system and has bigger information.


We would like to thank CNPq and CAPES for partial financial support. I would like to thank Moises Bravo Gaete and Jackson Levi Said for fruitful discussions at the end of this work.

%
%
%

%
%

\end{document}